\documentclass[
  11pt,
  letterpaper,
  reprint,
  nofootinbib,
  notitlepage,
  superscriptaddress,
  aps,prl
]{revtex4-2}

\usepackage{lineno}
\usepackage{amsmath,amssymb,amsfonts}
\usepackage{empheq}
\usepackage{physics}
\usepackage[title]{appendix}

\usepackage{amsthm}

\theoremstyle{remark}

\usepackage{subdepth}
\usepackage{xr}
\usepackage{bm}
\renewcommand{\mathbf}{\bm}
\usepackage{dsfont}
\renewcommand{\mathbb}{\mathds}

\usepackage[svgnames,dvipsnames]{xcolor}
\usepackage{graphicx}
\definecolor{NewBlue}{rgb}{0.1, 0.1, 0.7}
\definecolor{NewRed}{rgb}{0.7, 0.1, 0.1}
\usepackage[colorlinks,
  linkcolor=Maroon,
  citecolor=NewBlue,
urlcolor=NewRed]{hyperref}

\usepackage{cleveref}

\usepackage{empheq}
\newcommand{\eqdef}{\coloneqq}

\renewcommand{\t}[1]{\mathrm{#1}}

\newcommand{\LigoMIT}{LIGO Laboratory, Massachusetts Institute of Technology, Cambridge, MA 02139, USA}
\newcommand{\MechMIT}{Department of Mechanical Engineering, Massachusetts Institute of Technology, Cambridge, MA 02139, USA}

\begin{document}

\title{Noise-Resilient Quantum Metrology}

\author{Hudson A. Loughlin}
\email{hudsonl@mit.edu}
\affiliation{\LigoMIT}

\author{Melissa A. Guidry}
\affiliation{\LigoMIT}

\author{Jacques Ding}
\affiliation{\LigoMIT}
\affiliation{Laboratoire Astroparticule et Cosmologie, Universit\'e Paris Cit\'e, Paris, 75000, France}

\author{Masaya Ono}
\affiliation{\LigoMIT}
\affiliation{Department of Physics, The University of Tokyo, Bunkyo, Tokyo, 113-0033, Japan}

\author{Malo Le Gall}
\affiliation{\LigoMIT}

\author{Benjamin Lou}
\affiliation{\LigoMIT}

\author{Eric Oelker}
\affiliation{\LigoMIT}

\author{Xinghui Yin}
\affiliation{\LigoMIT}

\author{Vivishek Sudhir}
\affiliation{\LigoMIT}
\affiliation{\MechMIT}

\author{Nergis Mavalvala}
\affiliation{\LigoMIT}

\date{\today}

\begin{abstract}
  Quantum metrology seeks to leverage the richness of quantum systems for making better measurements than are possible using only classical resources in order to gain a ``quantum advantage''. Quantum metrology schemes must also be resilient against noise to be useful in practice. Simultaneously achieving quantum advantage and noise resilience requires an end-to-end analysis of quantum measurement schemes to assess their theoretical sensitivity, feasibility, and noise robustness. We demonstrate this approach through the development of a novel optical interferometer based on squeezed vacuum light. We propose a scheme that relies on a nonlinear phase estimation procedure, which allows us to shift the frequency of noise away from the signal band, resulting in a high degree of noise resilience. This enables us to achieve sensitivity with Heisenberg scaling in the lossless limit and sensitivity below the standard quantum limit (SQL) in practice. It also enables the first experimental demonstration of quantum-optimal Bayesian signal estimation in a balanced interferometer. We expect this end-to-end design approach to enable the development of a variety of useful quantum measurement protocols going forward.
\end{abstract}

\maketitle


The promise of quantum metrology \cite{Hel76,giovannetti2011,Barbieri22} is that
expensive quantum resources can be used economically to make better measurements than
are possible using only classical resources. 
Quantum metrology techniques \cite{DEMKOWICZDOBRZANSKI2015345,braun2018rev} that only focus on measurement sensitivity, without regard to noise resilience, are of limited practical utility.
One solution is quantum error correction for metrology
 \cite{Unden16,reiter2017,wang2022,ni2025}. However, that is
resource-expensive, and viable schemes for error-corrected optical metrology have remained elusive.

We propose and demonstrate an end-to-end quantum metrology design paradigm
that can achieve Heisenberg scaling --- the ultimate sensitivity bound allowed by
quantum mechanics --- in a continuous-wave (CW) optical interferometer 
while simultaneously being noise-resilient.
The end-to-end design paradigm optimizes the totality of the metrology task by considering the
quantum state used to probe the quantity of interest, the measurement of the resulting
state, and the subsequent estimation of the quantity of interest from the measurement record in the system's design.

The resulting scheme uses a pair of CW squeezed vacua to probe the phase signal transduced by 
a balanced interferometer; the output of the interferometer is measured by a pair of homodyne detectors; and, importantly, a nonlinear estimator is used to infer the phase signal. Using a Bayesian version of the estimator, we can also infer arbitrary waveform signals. This scheme is provably resource-efficient in terms of the number of photons required to achieve a given sensitivity for waveform estimation, approaching Heisenberg scaling, while simultaneously being noise-resilient. To our knowledge, this is the first CW interferometer with sensitivity approaching Heisenberg scaling for full waveform estimation.

\begin{figure*}[t]
  \centering
  \includegraphics[width=1.0\textwidth]{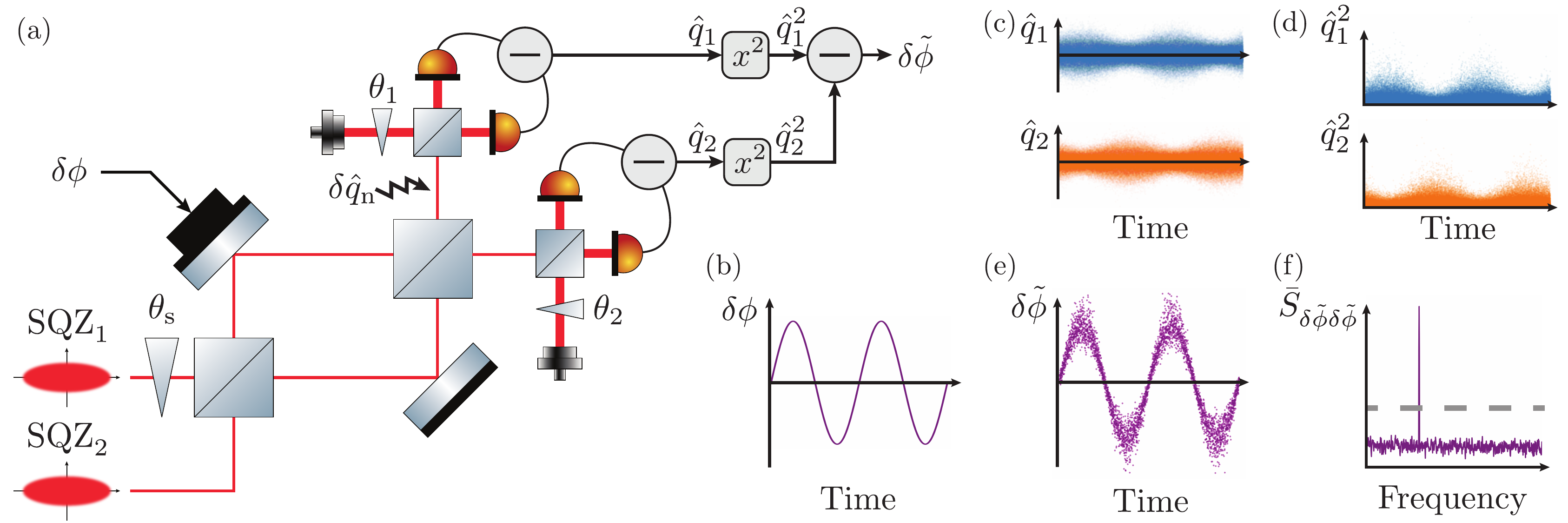}
  \caption{\label{fig:mziSchematic} \textbf{Schematic of the squeezed light interferometer and phase estimator.}
    \textbf{(a)} A pair of CW squeezed vacuum states, with relative phase $\theta_s$, are used to sense a phase signal, $\delta \phi(t)$, injected
    into an arm of a Mach-Zehnder interferometer. A pair of homodyne detectors measure the quadratures of the output field corresponding to
    the angles $\theta_{1,2}$. Each homodyne detector produces a record from which the phase signal is estimated by squaring the records and subtracting them.
    \textbf{(b)} The sinusoidal phase signal injected into the interferometer.
    \textbf{(c)} The resulting homodyne records are zero-mean with variances that oscillate like the phase signal, but out of phase.
    \textbf{(d,e)} The principle of our nonlinear estimator.
    \textbf{(d)} The rectified homodyne records, whose means oscillate out of phase.
    \textbf{(e)} Subtracting the rectified records recovers the injected phase signal.
    \textbf{(f)} The PSD of the estimator shows the phase signal as a spike at its frequency, riding atop white noise.
  This white noise is below the SQL (gray dashed line), exhibits Heisenberg scaling with photon flux, and saturates the spectral QCRB.}
\end{figure*}

\subsection{End-to-End Design}

The goal of quantum metrology is the inference of physical quantities at the 
limits imposed by quantum mechanics \cite{Hel76,giovannetti2006metrology_prl,Barbieri22}.
A typical scenario is the estimation of a phase shift $\phi$ transduced onto a quantum state $\hat \rho_\phi$; the state is measured and an estimator $\tilde{\phi}$ is constructed from the measurement outcomes.
For an unbiased estimator ($\langle\tilde{\phi}\rangle = \phi$), its precision is bounded by the quantum Cram\'er-Rao bound (QCRB) \cite{Hel76,Braunstein_geo_of_states,Paris09,Demk15}: $\t{Var}[\tilde{\phi}] \geq 1/F_Q$, where $F_Q$ is the quantum Fisher information (QFI).
With $n$ average photons, $F_Q \propto n$ for unentangled states --- the standard quantum limit (SQL) --- while entangled states can achieve $F_Q \propto n^2$, so-called Heisenberg scaling.

We seek to perform CW phase estimation that (1) achieves Heisenberg scaling in principle and surpasses the SQL in practice, (2) is robust against environmental noise, and (3) is realizable with current technology.

Generically, achieving Heisenberg scaling requires noiseless transduction.
Among schemes that achieve Heisenberg scaling \cite{Yurke86,Holland93,Lee2002,Anisimov2010,Lang13,lang2014optimal,lou2024trade}, only those using coherent and/or squeezed 
vacuum states support CW operation.
Passive balanced interferometers are naturally immune to common-mode noises in contrast to
active interferometers \cite{Yurke86,Plick2010,Jing11,Hudelist14,Anderson17,Zuo20}. 
We therefore use a passive balanced (Mach-Zehnder) interferometer as the transducer.

For such a device with total input photon number $n$, the QFI satisfies $F_Q \leq n(n+2)$ \cite{lang2014optimal}, saturated when the input is a pair of squeezed states.
We find this scaling extends to continuous-time phase estimation with a pair of squeezed vacuum inputs (as in \cref{fig:mziSchematic}, see Supplementary Information).
Squeezing is realistically only ever available over a finite bandwidth $\Delta\Omega$ about a center frequency $\Omega_0$, i.e. the quadrature power spectral density (PSD) of each squeezed state is 
$\bar{S}_{xx}[\Omega] = (V_\pm - 1/2) H[\Omega] + 1/2$, where $V_\pm$ are the squeezing and anti-squeezing factors, $H[\Omega]$ is the squeezing bandwidth filter ($H=1$ within $\Delta\Omega/2$ of $\Omega_0$, zero otherwise), and $V_\pm = e^{\pm 2r}/2$ for a pure squeezed state (see Supplementary Information).
For pure squeezed inputs without loss, the precision of any CW phase estimator is bounded by the spectral QCRB (see Supplementary Information)
\begin{equation}\label{eq:SQCRBmain}
  \bar{S}_{\delta\tilde{\phi}\delta{\tilde\phi}}[\Omega] \geq \frac{1}{ \frac{\Delta\Omega}{\pi} n(n + 2)},
\end{equation}
for $\Omega \ll \Delta\Omega$, where $n \eqdef V_+ + V_- -1$ is the photon number occupancy per mode.
For $n\gg 2$, this gives Heisenberg scaling: $\bar{S}_{\delta\tilde{\phi}\delta{\tilde\phi}}^{1/2} \sim 1/n$.

\begin{figure*}[t]
    \includegraphics[width=1.0\textwidth]{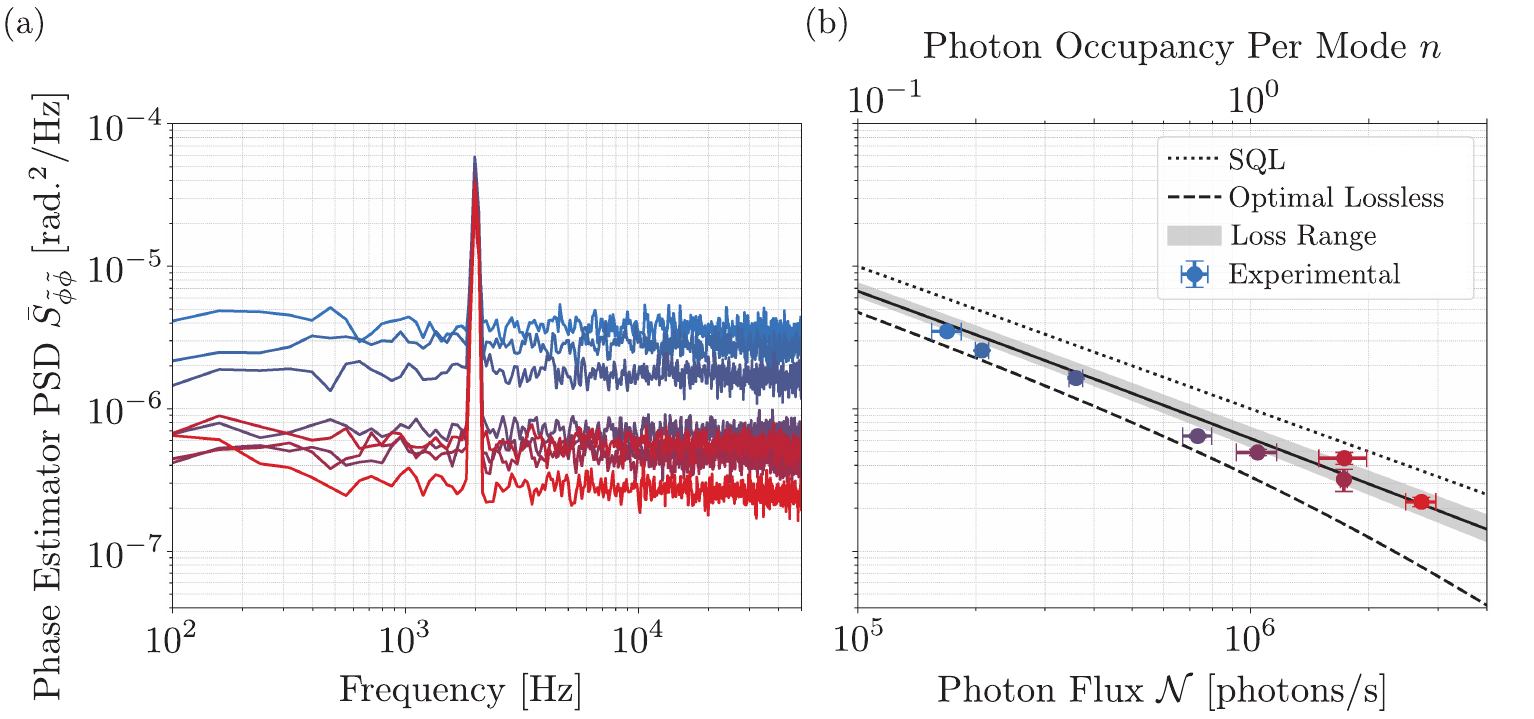}
    \centering
    \caption{\textbf{Phase precision vs photon flux.}
      \textbf{(a)} PSD of the phase estimator as photon flux through the interferometer is increased.
      The injected phase signal at $2$~kHz is visible above the quantum noise-limited noise floor.
      \textbf{(b)} Phase precision (with the same y-axis as panel (a)) taken from the mean value of the traces in panel (a) above 2.8 kHz, as a function of photon flux,
      extracted from state tomography measurements.
      Top axis shows the equivalent photon occupancy per mode.
      When $n \ll 2$, the optimal scaling is $1/(2n)$. When $n \gg 2$, the optimal scaling is $1/n^2$, Heisenberg scaling.
      The error bars in the photon flux are estimated from measurements taken before and after each PSD measurement.
      The error bars in the phase precision are 1-sigma standard deviations from a series of five PSD measurements
    and arise from system drifts, rather than from PSD estimation uncertainty. The solid black line shows the theoretical phase precision with the mean level of measured loss.}
    \label{fig:phaseSensitivity}
  \end{figure*}

Finally, a noise-resilient measurement and estimation scheme is required to achieve this scaling in
practice. We extend the quasi-static result of Ref.~\cite{Nielsen23} to the CW setting, and show
the homodyne records, filtered to pass the squeezing band, $\tilde{q}_i^\t{out}(t) \eqdef
\int H(t-t^\prime) \hat{q}_i^\text{out}(t^\prime) \t{d}t'$, and fed into the (unbiased) 
maximum likelihood estimator (MLE)
\begin{equation}\label{eq:optEst}
  \delta\tilde{\phi}(t) = \frac{\tilde{q}_1^\text{out}(t)^2 - \tilde{q}_2^\text{out}(t)^2}
  {(V_+ - V_-) \Delta\Omega/\pi},
\end{equation}
has a PSD
\begin{equation}\label{eq:estPsdLossy}
  \bar{S}_{\delta\tilde{\phi}\delta{\tilde\phi}}[\Omega] = \frac{1+2 (1-\eta^2) n}{ \frac{\Delta\Omega}{\pi} n (n+2 \eta^2 )},
\end{equation}
which saturates the spectral QCRB [\cref{eq:SQCRBmain}] when the loss $1-\eta^2 \rightarrow 0$.
The quadratic nonlinearity rectifies the zero-mean homodyne records, whose variance encodes the phase signal; rectifying and subtracting the two outputs recovers it (\cref{fig:mziSchematic}(c-f)).
Thus, homodyne detection with the estimator in \cref{eq:optEst} is optimal: no quantum measurement and estimation scheme can achieve better phase precision.

\subsection{Approaching Heisenberg Scaling}

We realize the interferometer by sending the fields from a pair of squeezed vacuum sources in phase ($\theta_\text{s} = 0$) into the inputs of a Mach-Zehnder interferometer, which is operated mid-fringe, and read out using a pair of homodyne detectors operating in quadrature ($\theta_1 = 0$ and $\theta_2 = \pi/2$). We observe squeezing in the band centered at 
$\Omega_0 = 2\pi\cdot450\,\t{kHz}$ with bandwidth $\Delta\Omega = 2\pi\cdot \t{500}\, \t{kHz}$; the interferometer is sensitive to signals up to $\Omega =
2\pi\cdot 50\, \t{kHz}$ (such that $\Omega \ll \Delta\Omega$). 

\begin{figure*}[t]
  \includegraphics[width=1.0\textwidth]{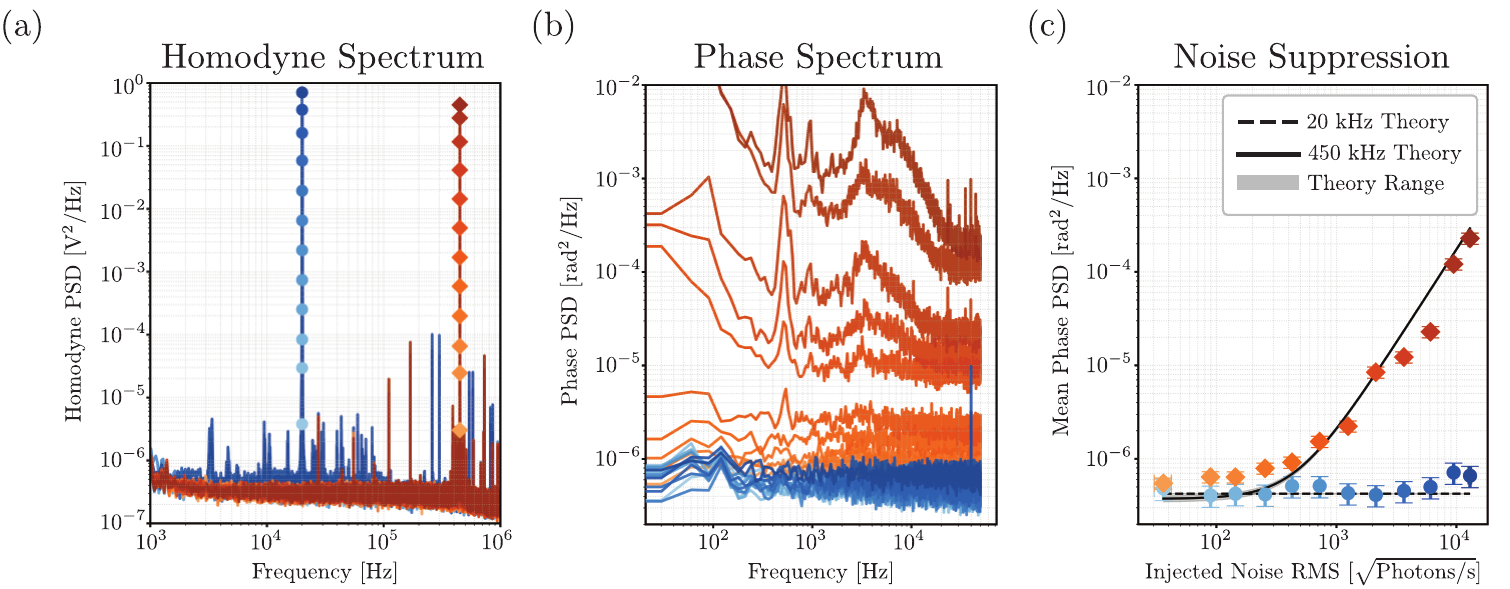}
  \centering
  \caption{\textbf{Noise characterization.}
    \textbf{(a)} Homodyne spectra for different injected noise levels. The dots indicate the peak PSD amplitudes.
    \textbf{(b)} The estimated phase spectra corresponding to the homodyne traces of the same color. The estimated phase spectra are not altered by the injection of a tone at 20 kHz, but increase significantly when a tone of the same amplitude is injected at 450 kHz.
    \textbf{(c)} The mean phase estimator PSD from 20-50 kHz from the traces in (b) as a function of the injected noise in units of the amplitude quadrature, $\hat{q}$. We see that the estimated phase PSD is insensitive to noise at 20 kHz even when a tone of the same amplitude injected at 450 kHz causes a large excess in the estimator's PSD.
    }
    \label{fig:calibration}
  \end{figure*}

In order to characterize the precision with which the phase signal can be estimated, and in particular its scaling with photon flux, we measure a series of PSDs of the phase estimator as the input squeezing level is varied.
Before and after each measurement, we perform state tomography to determine the photon flux and losses.
In each measurement, we inject a calibration tone into the interferometer's differential phase, $\delta\phi(t)$, at 2 kHz, which allows us to verify that the system is at its optimal operating point (see Supplementary Information).
\Cref{fig:phaseSensitivity}(a) shows the PSD measurement at the optimal operating point as the input squeezing level is varied.
From these PSD measurements, we compute the imprecision in the phase estimator as the mean value of the PSD
between 2.8 and 40~kHz (which excludes the 2~kHz calibration tone and any low-frequency noise).

\Cref{fig:phaseSensitivity}(b) shows the observed scaling of the phase precision with photon flux.
The solid black line (gray band) shows the model of the expected scaling of the precision with photon flux
for the mean optical loss (range of measured losses), and the dashed black line shows the model of the expected phase precision in the absence of losses, which achieves Heisenberg scaling. These theoretical curves are determined from \cref{eq:estPsdLossy} with experimentally measured losses and no free parameters.
The dotted line shows the SQL; the phase precision of any measurement using only classical means must lie on or above this line. All our measurements lie beneath the SQL and approach Heisenberg scaling.

\subsection{Noise Resilience}

To investigate the interferometer's robustness against noises with frequencies in the signal band, we intentionally inject noise between one interferometer output and the associated homodyne detector ($\delta \hat q_\text{n}$ in \cref{fig:mziSchematic}). We expect that noise outside of the measured squeezing band has no effect on the measurement's noise floor, quantified by the PSD of the phase estimator. Indeed, the blue traces in \cref{fig:calibration}(a,b) show that injecting a noise tone at 20 kHz has no effect on the estimator's noise, since the nonlinear estimator shifts this noise out of the signal band. For any noise tone outside of the squeezing band, we expect that the phase noise level is given by \cref{eq:estPsdLossy} regardless of the amplitude of the injected tone. This decoupling between signal and noise frequencies is 
characteristic of a nonlinear estimator.
  
In contrast, injecting a noise tone of the same amplitude at 450 kHz, in the squeezing band, does reduce phase sensitivity, shown in the red traces in \cref{fig:calibration}(a,b). 
This noise is spectrally flat for relatively low injected modulation tones but for larger tones begins to show additional excess noise below about 20 kHz due to impurities in the injected tone (also visible on the homodyne signal in \cref{fig:calibration}(a)). Quantitatively, injecting a single-frequency noise tone with root-mean-squared amplitude $q_\text{rms}$ inside the squeezing band leads to the estimated phase noise 
(see Supplementary Information for details)
\begin{equation}\label{eq:estPsdNoiseInjection}
  \bar S_{\delta \tilde \phi \delta \tilde \phi}^\text{noise} [\Omega] = \left( 1 + \frac{8\pi (n+1)q_\text{rms}^2}{(1+2(1-\eta^2)n) \Delta \Omega} \right) \bar S_{\delta \tilde \phi \delta \tilde \phi} [\Omega].
\end{equation}
\Cref{fig:calibration}(c) compares the measured phase noise 
for a noise tone 
at 20 or 450 kHz, corresponding to the models in \cref{eq:estPsdLossy} and \cref{eq:estPsdNoiseInjection}, respectively. 
In both cases, the photon number and loss are determined from data collected as in \cref{fig:calibration}(a) and the amplitude of the injected tone is calibrated in quadrature units by referencing against shot noise. There are no free parameters. In both cases the measured noise floor agrees with theoretical expectations in this spectral region. This confirms that our interferometer achieves a high degree of low-frequency noise resilience, which is due to the nonlinear estimator.

\begin{figure*}[t]
\centering
\includegraphics[width=1.0\textwidth]{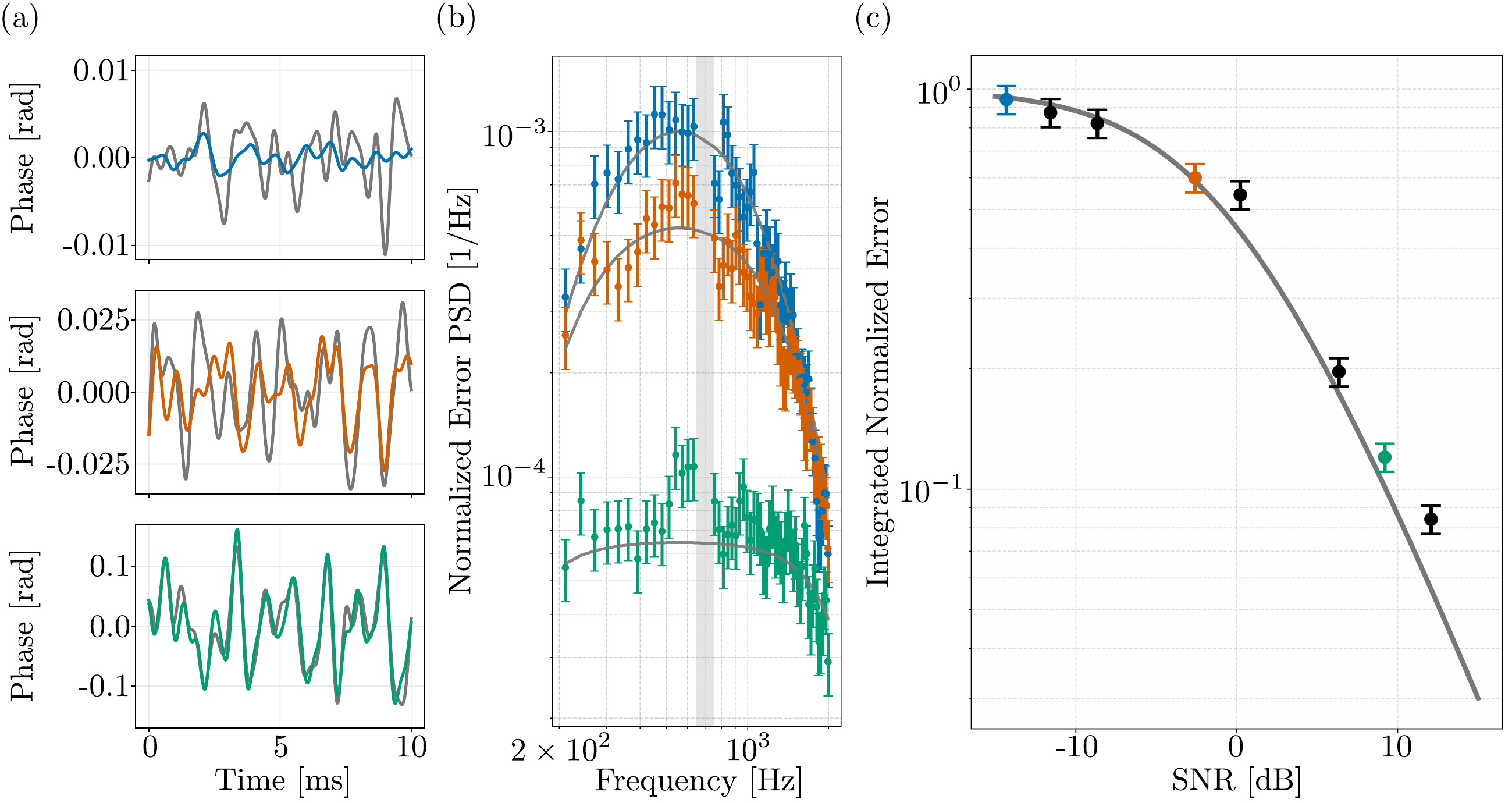}
    \caption{\label{fig:waveformEstimation} \textbf{Waveform estimation at the quantum noise limit.} \textbf{(a)} Timeseries waveform estimation for low, intermediate, and high SNRs. The gray traces are the true bandpassed signals and the colored traces are the estimated signals. \textbf{(b)} The corresponding normalized error PSDs for the time traces of the same color, with one-sigma error bars. The shaded region from 650 to 750 Hz contains a sharp piezo feature and was excluded from the timeseries and error analysis (see Supplementary Information). The gray traces are the expected error PSDs from our parameter-free model. \textbf{(c)} The integrated error over the signal band with three-sigma error bars compared with our parameter-free model. The black data points correspond to data not plotted in (a) and (b) but collected in the same manner. We see that in all cases we are able to estimate the injected waveform as accurately as quantum mechanics allows.}
\end{figure*}

\subsection{Waveform Estimation}

Typical applications of interferometry involve the estimation of a random 
time-varying phase signal.
Suppose a random waveform $\delta\phi(t)$ is transduced by the interferometer and is estimated by 
an estimator $\delta\tilde\phi^\text{wave}(t)$. 
Assuming that we know the statistics of the waveform (but not the waveform itself), say the prior 
PSD $\bar{S}^\text{prior}_{\delta\phi \delta\phi}$, the waveform can be estimated with a precision
bounded by the waveform CRB \cite{van1968,trees2007,Tsang11}: 
\begin{equation}\label{eq:waveformQcrb}
    \bar{S}_{\Delta\phi \Delta\phi}[\Omega] \geq \frac{\bar{S}^\text{prior}_{\delta\phi \delta\phi}[\Omega] \bar{S}_{\delta\tilde\phi \delta\tilde\phi}[\Omega]}{\bar{S}_{\delta\tilde\phi \delta\tilde\phi}[\Omega] + \bar{S}^\text{prior}_{\delta\phi \delta\phi}[\Omega]},
\end{equation}
where $\Delta\phi$ is the estimation error.
The bound can be saturated by smoothing the estimator $\delta\tilde\phi^\text{wave}$ with a (non-causal) 
Wiener filter with frequency response $\bar{S}^\text{prior}_{\delta\phi \delta\phi}[\Omega]/(\bar{S}^\text{prior}_{\delta\phi \delta\phi}[\Omega] + \bar{S}_{\delta\tilde\phi \delta\tilde\phi}[\Omega])$ \cite{Iwasawa13}. 
Importantly, the waveform CRB in \cref{eq:waveformQcrb} is 
a Bayesian bound, which exhibits Heisenberg scaling only when the measurement  
conveys more information than the prior, in which case $\bar{S}_{\delta\tilde\phi \delta\tilde\phi}[\Omega] \ll \bar{S}^\text{prior}_{\delta\phi \delta\phi}[\Omega]$ such that $\bar{S}_{\Delta\phi \Delta\phi}[\Omega] \gtrsim \bar{S}_{\delta\tilde\phi \delta\tilde\phi}[\Omega]$.

Experimentally, we synthesize a random waveform and use it to drive the interferometer's differential phase. 
The waveform is then inferred by Wiener filtering the output of the nonlinear phase estimator.
\Cref{fig:waveformEstimation}(a) shows the first 10 ms of the bandpassed timeseries of three injected waveforms and the estimated waveforms from the processed homodyne records. 
The SNR of the waveform is increased successively from top to bottom, which leads to successively better waveform reconstruction. The middle column shows the resulting waveform estimation error normalized by the integrated spectrum of the injected waveform, compared with the optimal value from \cref{eq:waveformQcrb} with the estimation error PSD estimated from the entire 1.06 second dataset. Each color corresponds to the dataset with the same color in \cref{fig:waveformEstimation}(a,b).

\Cref{fig:waveformEstimation}(c) shows the integrated normalized estimation error compared with the bound from  \cref{eq:waveformQcrb}. The colored data points correspond to those shown in \cref{fig:waveformEstimation}(a) and (b) and the black data points were taken in the same manner but the corresponding data are not shown in \cref{fig:waveformEstimation}(a,b).

We see that the data are consistent with the optimal waveform estimation bound.
Furthermore, we see that the measured data tracks the optimal bound from the low SNR regime to the high SNR regime. This demonstrates that the noise resilience of our system allows it to reconstruct phase signals in the audio band, here between 200 Hz and 2 kHz, at the waveform CRB limit imposed by quantum noise.

\subsection{Discussion}

By applying an end-to-end design approach, we have demonstrated an optical interferometer that suppresses common noise sources, achieves sensitivity below the SQL and approaching Heisenberg scaling, and realizes optimal waveform estimation. The observed scaling is consistent with our theoretical model [\cref{eq:estPsdLossy}] given measured losses of $20$--$35\%$; reducing these will allow arbitrarily close approach to Heisenberg scaling in CW interferometry.

A key technical advantage of our nonlinear estimator is that squeezing at higher frequencies is down-mixed to lower frequencies, so, unlike linear measurements, the band of optimal phase precision need not coincide with the band of quantum-noise-limited homodyne detection. This enables quantum-limited phase precision below 100~Hz (\cref{fig:calibration}(b)) without mitigating back-scatter and other low-frequency noise sources.

Using this noise-resilient interferometric measurement scheme, we demonstrated, for the first time, estimation of a time-varying optical phase at the quantum noise-limited waveform CRB in a balanced interferometer, illustrating the passage of quantum estimation from the prior-dominated to measurement-dominated regime \cite{Iwasawa13}.

Achieving Heisenberg scaling and optimal waveform estimation requires end-to-end design of the interferometer's input quantum states with the measurement and estimation procedure: the input must permit Heisenberg scaling according to the spectral QCRB, and this scaling must be attainable by a practical measurement and estimator. This end-to-end design principle will be essential for future quantum-enhanced sensors, with applications wherever small phase shifts must be measured at minimal optical power, including biological imaging \cite{park2018quantitative,CasBow21,laissue2017assessing,icha2017phototoxicity} and atomic-state readout in quantum computers \cite{Wang25}.

\bibliography{ref_squeezed_state_interferometry}

@incollection{DEMKOWICZDOBRZANSKI2015345,
title = {Chapter Four - Quantum Limits in Optical Interferometry},
booktitle = {Progress in Optics},
editor = {E. Wolf},
publisher = {Elsevier},
volume = {60},
pages = {345-435},
year = {2015},
issn = {0079-6638},
doi = {https://doi.org/10.1016/bs.po.2015.02.003},
url = {https://www.sciencedirect.com/science/article/pii/S0079663815000049},
author = {Rafal Demkowicz-Dobrzański and Marcin Jarzyna and Jan Kołodyński},
}

@book{trees2007,
  title={Bayesian bounds for parameter estimation and nonlinear filtering/tracking},
  author={Trees, Harry L Van and Bell, Kristine L},
  year={2007},
  publisher={Wiley-IEEE press}
}

@article{Zuo20,
  title = {Quantum Interferometer Combining Squeezing and Parametric Amplification},
  author = {Zuo, Xiaojie and Yan, Zhihui and Feng, Yanni and Ma, Jingxu and Jia, Xiaojun and Xie, Changde and Peng, Kunchi},
  journal = {Phys. Rev. Lett.},
  volume = {124},
  issue = {17},
  pages = {173602},
  numpages = {6},
  year = {2020},
  month = {May},
  publisher = {American Physical Society},
  doi = {10.1103/PhysRevLett.124.173602},
  url = {https://link.aps.org/doi/10.1103/PhysRevLett.124.173602}
}

@article{Anderson17,
	author = {Brian E. Anderson and Prasoon Gupta and Bonnie L. Schmittberger and Travis Horrom and Carla Hermann-Avigliano and Kevin M. Jones and Paul D. Lett},
	doi = {10.1364/OPTICA.4.000752},
	journal = {Optica},
	month = {Jul},
	number = {7},
	pages = {752--756},
	publisher = {Optica Publishing Group},
	title = {Phase sensing beyond the standard quantum limit with a variation on the SU(1,1) interferometer},
	url = {https://opg.optica.org/optica/abstract.cfm?URI=optica-4-7-752},
	volume = {4},
	year = {2017}}

@article{Hudelist14,
	author = {Hudelist, F. and Kong, Jia and Liu, Cunjin and Jing, Jietai and Ou, Z. Y. and Zhang, Weiping},
	date = {2014/01/29},
	doi = {10.1038/ncomms4049},
	id = {Hudelist2014},
	isbn = {2041-1723},
	journal = {Nature Communications},
	number = {1},
	pages = {3049},
	title = {Quantum metrology with parametric amplifier-based photon correlation interferometers},
	url = {https://doi.org/10.1038/ncomms4049},
	volume = {5},
	year = {2014}}

@article{Jing11,
	author = {Jing, Jietai and Liu, Cunjin and Zhou, Zhifan and Ou, Z. Y. and Zhang, Weiping},
	doi = {10.1063/1.3606549},
	issn = {0003-6951},
	journal = {Applied Physics Letters},
	month = {07},
	number = {1},
	pages = {011110},
	title = {Realization of a nonlinear interferometer with parametric amplifiers},
	url = {https://doi.org/10.1063/1.3606549},
	volume = {99},
	year = {2011}}

@article{Plick2010,
doi = {10.1088/1367-2630/12/8/083014},
url = {https://doi.org/10.1088/1367-2630/12/8/083014},
year = {2010},
month = {aug},
publisher = {},
volume = {12},
number = {8},
pages = {083014},
author = {Plick, William N and Dowling, Jonathan P and Agarwal, Girish S},
title = {Coherent-light-boosted, sub-shot noise, quantum interferometry},
journal = {New Journal of Physics}
}

@article{Iwasawa13,
  title = {Quantum-Limited Mirror-Motion Estimation},
  author = {Iwasawa, Kohjiro and Makino, Kenzo and Yonezawa, Hidehiro and Tsang, Mankei and Davidovic, Aleksandar and Huntington, Elanor and Furusawa, Akira},
  journal = {Phys. Rev. Lett.},
  volume = {111},
  issue = {16},
  pages = {163602},
  numpages = {5},
  year = {2013},
  month = {Oct},
  publisher = {American Physical Society},
  doi = {10.1103/PhysRevLett.111.163602},
  url = {https://link.aps.org/doi/10.1103/PhysRevLett.111.163602}
}

@article{Lee2002,
author = {Hwang Lee and Pieter Kok and Jonathan P. Dowling},
title = {A quantum Rosetta stone for interferometry},
journal = {Journal of Modern Optics},
volume = {49},
number = {14-15},
pages = {2325--2338},
year = {2002},
publisher = {Taylor \& Francis},
doi = {10.1080/0950034021000011536},
URL = {https://doi.org/10.1080/0950034021000011536},
eprint = {https://doi.org/10.1080/0950034021000011536}
}

@article{wang2022,
	title = {Quantum-enhanced radiometry via approximate quantum error correction},
	volume = {13},
	issn = {2041-1723},
	url = {https://doi.org/10.1038/s41467-022-30410-8},
	doi = {10.1038/s41467-022-30410-8},
	number = {1},
	journal = {Nature Communications},
	author = {Wang, W. and Chen, Z.-J. and Liu, X. and Cai, W. and Ma, Y. and Mu, X. and Pan, X. and Hua, Z. and Hu, L. and Xu, Y. and Wang, H. and Song, Y. P. and Zou, X.-B. and Zou, C.-L. and Sun, L.},
	month = jun,
	year = {2022},
	pages = {3214},
}

@article{ni2025,
  title={Autonomous quantum error correction beyond break-even and its metrological application},
  author={Ni, Zhongchu and Hu, Ling and Cai, Yanyan and Zhang, Libo and Mai, Jiasheng and Deng, Xiaowei and Zheng, Pan and Liu, Song and Zheng, Shi-Biao and Xu, Yuan and others},
  journal={arXiv preprint arXiv:2509.26042},
  year={2025},
  url={https://arxiv.org/abs/2509.26042}
}

@article{reiter2017,
	title = {Dissipative quantum error correction and application to quantum sensing with trapped ions},
	volume = {8},
	issn = {2041-1723},
	url = {https://doi.org/10.1038/s41467-017-01895-5},
	doi = {10.1038/s41467-017-01895-5},
	number = {1},
	journal = {Nature Communications},
	author = {Reiter, F. and Sørensen, A. S. and Zoller, P. and Muschik, C. A.},
	month = nov,
	year = {2017},
	pages = {1822},
}

@article{Unden16,
  title = {Quantum Metrology Enhanced by Repetitive Quantum Error Correction},
  author = {Unden, Thomas and Balasubramanian, Priya and Louzon, Daniel and Vinkler, Yuval and Plenio, Martin B. and Markham, Matthew and Twitchen, Daniel and Stacey, Alastair and Lovchinsky, Igor and Sushkov, Alexander O. and Lukin, Mikhail D. and Retzker, Alex and Naydenov, Boris and McGuinness, Liam P. and Jelezko, Fedor},
  journal = {Phys. Rev. Lett.},
  volume = {116},
  issue = {23},
  pages = {230502},
  numpages = {6},
  year = {2016},
  month = {Jun},
  publisher = {American Physical Society},
  doi = {10.1103/PhysRevLett.116.230502},
  url = {https://link.aps.org/doi/10.1103/PhysRevLett.116.230502}
}

@article{giovannetti2011,
	title = {Advances in quantum metrology},
	volume = {5},
	issn = {1749-4893},
	url = {https://doi.org/10.1038/nphoton.2011.35},
	doi = {10.1038/nphoton.2011.35},
	number = {4},
	journal = {Nature Photonics},
	author = {Giovannetti, Vittorio and Lloyd, Seth and Maccone, Lorenzo},
	month = apr,
	year = {2011},
	pages = {222--229},
}

@book{van1968,
  title={Detection, Estimation, and Modulation Theory Part 1: Detection Estimation, and Linear Modulation Theory},
  author={Van Trees, Harry L},
  year={1968},
  publisher={Wiley, New York}
}

@article{CasBow21,
	author = {Casacio, Catxere A. and Madsen, Lars S. and Terrasson, Alex and Waleed, Muhammad and Barnscheidt, Kai and Hage, Boris and Taylor, Michael A. and Bowen, Warwick P.},
	doi = {10.1038/s41586-021-03528-w},
	journal = {Nature},
	number = {7862},
	pages = {201},
	title = {Quantum-enhanced nonlinear microscopy},
	url = {https://www.nature.com/articles/s41586-021-03528-w},
	volume = {594},
	year = {2021}}

@article{Paris09,
	author = {Paris, Matteo},
	doi = {10.1142/S0219749909004839},
	journal = {International Journal of Quantum Information},
	pages = {125--137},
	title = {Quantum {Estimation} for {Quantum} {Technology}},
	url = {http://www.worldscientific.com/doi/abs/10.1142/S0219749909004839},
	volume = {07},
	year = {2009}}

@incollection{Demk15,
	author = {Demkowicz-Dobrza{\'n}ski, Rafal and Jarzyna, Marcin and Ko{\l}ody{\'n}ski, Jan},
	booktitle = {Progress in {Optics}},
	chapter = {4},
	doi = {10.1016/bs.po.2015.02.003},
	editor = {Wolf, E.},
	pages = {345--435},
	publisher = {Elsevier},
	title = {{Quantum} {Limits} in {Optical} {Interferometry}},
	url = {https://www.sciencedirect.com/science/article/pii/S0079663815000049},
	volume = {60},
	year = {2015}}

@book{Hel76,
	author = {Helstrom, Carl W},
	publisher = {Academic Press},
	title = {{Quantum Detection and Estimation Theory}},
	year = {1976}}

@article{Wang25,
	author = {Wang, Jian and Huang, Dong-Yu and Zhou, Xiao-Long and Shen, Ze-Min and He, Si-Jian and Huang, Qi-Yang and Liu, Yi-Jia and Li, Chuan-Feng and Guo, Guang-Can},
	doi = {10.1103/PhysRevLett.134.240802},
	issue = {24},
	journal = {Phys. Rev. Lett.},
	month = {Jun},
	numpages = {7},
	pages = {240802},
	publisher = {American Physical Society},
	title = {Ultrafast High-Fidelity State Readout of Single Neutral Atom},
	url = {https://link.aps.org/doi/10.1103/PhysRevLett.134.240802},
	volume = {134},
	year = {2025}}

@article{park2018quantitative,
	author = {Park, YongKeun and Depeursinge, Christian and Popescu, Gabriel},
	journal = {Nature Photonics},
	number = {10},
	pages = {578--589},
	title = {Quantitative phase imaging in biomedicine},
	url = {https://www.nature.com/articles/s41566-018-0253-x},
	volume = {12},
	year = {2018}}

@article{icha2017phototoxicity,
	author = {Icha, Jaroslav and Weber, Michael and Waters, Jennifer C and Norden, Caren},
	journal = {BioEssays},
	number = {8},
	pages = {1700003},
	publisher = {Wiley Online Library},
	title = {Phototoxicity in live fluorescence microscopy, and how to avoid it},
	url = {https://onlinelibrary.wiley.com/doi/full/10.1002/bies.201700003},
	volume = {39},
	year = {2017}}

@article{laissue2017assessing,
	author = {Laissue, P Philippe and Alghamdi, Rana A and Tomancak, Pavel and Reynaud, Emmanuel G and Shroff, Hari},
	journal = {Nature Methods},
	number = {7},
	pages = {657--661},
	title = {Assessing phototoxicity in live fluorescence imaging},
	url = {https://www.nature.com/articles/nmeth.4344},
	volume = {14},
	year = {2017}}

@article{Holland93,
	author = {Holland, M. J. and Burnett, K.},
	doi = {10.1103/PhysRevLett.71.1355},
	issue = {9},
	journal = {Phys. Rev. Lett.},
	month = {Aug},
	numpages = {0},
	pages = {1355--1358},
	publisher = {American Physical Society},
	title = {Interferometric detection of optical phase shifts at the Heisenberg limit},
	url = {https://link.aps.org/doi/10.1103/PhysRevLett.71.1355},
	volume = {71},
	year = {1993}}

@article{Nielsen23,
	author = {Nielsen, Jens A. H. and Neergaard-Nielsen, Jonas S. and Gehring, Tobias and Andersen, Ulrik L.},
	date = {2023/03/24/},
	day = {24},
	doi = {10.1103/PhysRevLett.130.123603},
	id = {10.1103/PhysRevLett.130.123603},
	j1 = {PRL},
	journal = {Physical Review Letters},
	journal1 = {Phys. Rev. Lett.},
	month = {03},
	number = {12},
	pages = {123603--},
	publisher = {American Physical Society},
	title = {Deterministic Quantum Phase Estimation beyond N00N States},
	url = {https://link.aps.org/doi/10.1103/PhysRevLett.130.123603},
	volume = {130},
	year = {2023}}

@article{lou2024trade,
	author = {Lou, Benjamin and Loughlin, Hudson A and Mavalvala, Nergis},
	journal = {arXiv preprint arXiv:2411.18832},
	title = {A Trade-Off Between Path Entanglement and Quantum Sensitivity},
	url = {https://arxiv.org/abs/2411.18832},
	year = {2024}}

@article{Tsang11,
	author = {Tsang, Mankei and Wiseman, Howard M. and Caves, Carlton M.},
	doi = {10.1103/PhysRevLett.106.090401},
	issue = {9},
	journal = {Phys. Rev. Lett.},
	month = {Mar},
	numpages = {4},
	pages = {090401},
	publisher = {American Physical Society},
	title = {Fundamental Quantum Limit to Waveform Estimation},
	url = {https://link.aps.org/doi/10.1103/PhysRevLett.106.090401},
	volume = {106},
	year = {2011}}

@article{lang2014optimal,
	author = {Lang, Matthias D and Caves, Carlton M},
	journal = {Physical Review A},
	number = {2},
	pages = {025802},
	publisher = {APS},
	title = {Optimal quantum-enhanced interferometry},
	volume = {90},
	year = {2014},
url={https://doi.org/10.1103/PhysRevA.90.025802}}

@article{Lang13,
	author = {Lang, Matthias D. and Caves, Carlton M.},
	doi = {10.1103/PhysRevLett.111.173601},
	issue = {17},
	journal = {Phys. Rev. Lett.},
	month = {Oct},
	numpages = {5},
	pages = {173601},
	publisher = {American Physical Society},
	title = {Optimal Quantum-Enhanced Interferometry Using a Laser Power Source},
	url = {https://link.aps.org/doi/10.1103/PhysRevLett.111.173601},
	volume = {111},
	year = {2013}}

@article{Anisimov2010,
	author = {Anisimov, Petr M. and Raterman, Gretchen M. and Chiruvelli, Aravind and Plick, William N. and Huver, Sean D. and Lee, Hwang and Dowling, Jonathan P.},
	doi = {10.1103/PhysRevLett.104.103602},
	issue = {10},
	journal = {Phys. Rev. Lett.},
	month = {Mar},
	numpages = {4},
	pages = {103602},
	publisher = {American Physical Society},
	title = {Quantum Metrology with Two-Mode Squeezed Vacuum: Parity Detection Beats the Heisenberg Limit},
	url = {https://link.aps.org/doi/10.1103/PhysRevLett.104.103602},
	volume = {104},
	year = {2010}}

@article{Yurke86,
	author = {Yurke, Bernard and McCall, Samuel L. and Klauder, John R.},
	doi = {10.1103/PhysRevA.33.4033},
	journal = {Phys. Rev. A},
	month = {Jun},
	pages = {4033--4054},
	title = {{SU(2)} and {SU(1,1)} interferometers},
	url = {https://link.aps.org/doi/10.1103/PhysRevA.33.4033},
	volume = {33},
	year = {1986}}

@article{Barbieri22,
	author = {Barbieri, Marco},
	doi = {10.1103/PRXQuantum.3.010202},
	issue = {1},
	journal = {PRX Quantum},
	month = {Jan},
	numpages = {24},
	pages = {010202},
	publisher = {American Physical Society},
	title = {Optical Quantum Metrology},
	url = {https://link.aps.org/doi/10.1103/PRXQuantum.3.010202},
	volume = {3},
	year = {2022}}

@article{giovannetti2006metrology_prl,
	author = {Giovannetti, Vittorio and Lloyd, Seth and Maccone, Lorenzo},
	journal = {Phys. Rev. Lett.},
	number = {1},
	pages = {010401},
	title = {Quantum metrology},
	url = {https://journals.aps.org/prl/abstract/10.1103/PhysRevLett.96.010401},
	volume = {96},
	year = {2006}}

@article{Braunstein_geo_of_states,
	author = {Braunstein, Samuel L. and Caves, Carlton M.},
	doi = {10.1103/PhysRevLett.72.3439},
	issue = {22},
	journal = {Phys. Rev. Lett.},
	month = {May},
	numpages = {0},
	pages = {3439--3443},
	publisher = {American Physical Society},
	title = {Statistical distance and the geometry of quantum states},
	url = {https://link.aps.org/doi/10.1103/PhysRevLett.72.3439},
	volume = {72},
	year = {1994}}

@article{braun2018rev,
	author = {Braun, Daniel and Adesso, Gerardo and Benatti, Fabio and Floreanini, Roberto and Marzolino, Ugo and Mitchell, Morgan W. and Pirandola, Stefano},
	doi = {10.1103/RevModPhys.90.035006},
	issue = {3},
	journal = {Rev. Mod. Phys.},
	month = {Sep},
	numpages = {47},
	pages = {035006},
	publisher = {American Physical Society},
	title = {Quantum-enhanced measurements without entanglement},
	url = {https://link.aps.org/doi/10.1103/RevModPhys.90.035006},
	volume = {90},
	year = {2018}}

\section{Acknowledgments}

We gratefully acknowledge the support of the National Science Foundation (NSF) through the LIGO operations cooperative agreement PHY-18671764464 and NSF award 2308969.
This work was made possible through the support of the Enrico Fermi Fellowships led by the Center for Spacetime and the Quantum, and supported by Grant ID \#63132 from the John Templeton Foundation. The opinions expressed in this publication are those of the author(s) and do not necessarily reflect the views of the John Templeton Foundation or those of the Center for Spacetime and the Quantum.
JD gratefully acknowledges the support of the EU Horizon 2020 Research and Innovation Program under the Marie Sklodowska-Curie Grant Agreement No. 101003460 (PROBES) and the MIT-France Seed Fund.
MO gratefully acknowledges the support of the Japan Society for the Promotion of Science (No. 202380256).
VS is partially supported by an NSF CAREER award (PHY-2441238).
HAL would like to acknowledge helpful conversations with M. Evans.
The authors thanks Sarah C. Benedict for copy editing the manuscript.
The authors thank Myron Macinnis, Mark Belanger, and the Edgerton machine shop for helpful equipment and advice.

\section{Author Contributions Statement}

BL and HAL carried out preliminary theory, which was significantly extended to a full time-dependent theory by JD and HAL under the advice of VS. HAL led the design and construction of the experiment, which was implemented by HAL, MO, MAG, JD, and MLG  under the advice of NM, VS, XY, and EO. HAL and MAG acquired and analyzed the data. HAL, JD, MAG, VS, and NM wrote the manuscript.


\end{document}